\documentclass[twocolappendix]{emulateapj}
\usepackage{amsmath}
\usepackage{subfigure}
\usepackage{graphicx}
\usepackage{color}
\usepackage{units}
\usepackage{lipsum}

\newcommand{\V}[1]{ \mathbf #1 }
\newcommand{\figext}[1]{#1.pdf}

\begin{document}

\title{Turbulent magnetic relaxation in pulsar wind nebulae}

\author{Jonathan Zrake} \affil{Kavli Institute for Particle Astrophysics and
  Cosmology, Stanford University, SLAC National Accelerator Laboratory, Menlo
  Park, CA 94025, USA}

\author{Jonathan Arons} \affil{Astronomy Department and
  Theoretical Astrophysics Center, University of California, Berkeley, 601
  Campbell Hall, Berkeley, CA 94720, USA}


\begin{abstract}

  We present a model for magnetic energy dissipation in a pulsar wind
  nebula. Better understanding of this process is required to assess the
  likelihood that certain astrophysical transients may be powered by the
  spin-down of a ``millisecond magnetar.'' Examples include superluminous
  supernovae, gamma-ray bursts, and anticipated electromagnetic counterparts to
  gravitational wave detections of binary neutron star coalescence. Our model
  leverages recent progress in the theory of turbulent magnetic relaxation to
  specify a dissipative closure of the stationary magnetohydrodynamic (MHD) wind
  equations, yielding predictions of the magnetic energy dissipation rate
  throughout the nebula. Synchrotron losses are treated self-consistently. To
  demonstrate the model's efficacy, we show that it can reproduce many features
  of the Crab Nebula, including its expansion speed, radiative efficiency, peak
  photon energy, and mean magnetic field strength. Unlike ideal MHD models of
  the Crab (which lead to the so-called $\sigma$-problem) our model accounts for
  the transition from ultra to weakly magnetized plasma flow, and for the
  associated heating of relativistic electrons. We discuss how the predicted
  heating rates may be utilized to improve upon models of particle transport and
  acceleration in pulsar wind nebulae. We also discuss implications for the Crab
  Nebula's $\gamma$-ray flares, and point out potential modifications to models
  of astrophysical transients invoking the spin-down of a millisecond magnetar.

\end{abstract}

\maketitle

\section{Introduction}
\label{sec:introduction}

A pulsar wind nebula (PWN) is a bubble of relativistic plasma energized by a
rapidly rotating, magnetized neutron star. The prototypical PWN is the Crab
Nebula, which is decisively the best studied celestial object beyond our solar
system, having served for decades as a testbed for theories of astrophysical
outflows and their radiative processes. PWNe are also of broad interest in
astroparticle physics, as potential sources of galactic positrons
\citep{Chi1996} and ultra-high energy cosmic rays \citep{Arons2002}. More
generally, PWNe have ultra-energetic (albeit hypothetical) counterparts in the
winds of so-called ``millisecond magnetars.'' Such exotic objects may be formed
in the coalescence of binary neutron star systems, in which case they could
yield the first electromagnetic counterparts to gravitational wave detections
\citep{Metzger2011, Zhang2013, Gao2013, Zrake2013, Metzger2014, Liu2016}. Or, if
formed during the core-collapse of a massive star, a millisecond magnetar could
re-energize the ejecta shell, helping to explain the light-curves of certain
hydrogen-poor superluminous supernovae \citep{Woosley2010, Kasen2010,
  Dessart2012, Metzger2013a, Kasen2016}.

Morphological and radiative characteristics of pulsar winds are shaped by the
rate with which they dissipate magnetic energy. This can be seen from basic
considerations of the Crab. While energy streams away from the pulsar in the
form of an ultra-magnetized plasma wind \citep[e.g.][]{Goldreich1969},
throughout the nebula it is shared equitably with particles
\citep[][]{Rees1974}. To be sure, low magnetization levels in the range of
$10^{-3}-10^{-2}$ are required to explain the nebula's expansion speed
\citep{Kennel1984}, synchrotron spectrum \citep{Kennel1984a}, and mildly prolate
appearance \citep{Begelman1992}. Such low magnetization cannot be attained in
the absence of dissipative effects \citep{Michel1969, Goldreich1970, Chiueh1991,
  Begelman1994}.

At the time of the early models, little was known about how and where magnetic
dissipation operates in a PWN, so the issue was dealt with
pragmatically. \cite{Kennel1984} assigned a small, nominal magnetization to the
plasma emerging from the inner nebula, and modeled its flow henceforth
adiabatically. By construction, this procedure leaves any dissipative processes,
and thus the electron heating profile unspecified; all the nebula's relativistic
electrons appear in this description to be sourced from the immediate vicinity
of the wind termination shock.

Today the details of magnetic energy dissipation in the Crab are much better
understood. Magnetic field supplied by the pulsar has distinct AC and DC
components, which dissipate in different ways. The AC field \citep[also referred
  to as ``striped wind'',][]{Michel1971, Coroniti1990, Michel1994} consists of
magnetic reversals, which dissipate by mixing and annihilating one another. The
striped wind fills the equatorial wedge of the freely expanding pulsar wind,
whose opening angle is determined by the pulsar's magnetic obliquity
\citep[e.g.][]{Komissarov2012}.  The pulsar wind's DC power on the other hand,
is carried outward by a regular toroidal magnetic field. After being decelerated
at the so-called wind termination shock \citep[or a series of
  shocks,][]{Lyubarsky2003a}, the toroidal field's hoop stress induces flow
toward the polar axis \citep{Lyubarsky2002}, yielding an unstable $z$-pinch
configuration \citep{Begelman1998} seen in X-ray observations
\citep[e.g.][]{Hester2002} as the Crab's polar jet feature. Dynamical
instabilities act continuously to convert the energy of freshly supplied
magnetic hoops into turbulence and eventually heat \citep{Porth2013a,
  Porth2013}.

Our aim in this paper is to develop a model for the flow of magnetized plasma in
a PWN, which captures these dissipative processes as they influence the object's
morphological and radiative characteristics. Most of the new ideas presented
here go toward crafting a physically motivated prescription for the dissipation
of the magnetic field by \emph{turbulent relaxation}. Our key assumption is that
relaxation proceeds dynamically, in the sense that magnetic energy density may
be assigned a local half-life $\tau$, determined at each point by a comoving
Alfv{\'e}n speed and ``eddy'' scale. This assumption is motivated by recent
numerical work, which reveals a tendency for magnetic field configurations to
relax and dissipate by exciting turbulent motions, provided that states of lower
magnetic energy are topologically accessible in the sense of
\cite{Taylor1974}. This principle was illustrated by \cite{East2015} and
\cite{Zrake2016} using high resolution numerical simulations of prototypical
force-free equilibria \citep[so-called ``ABC'' fields,][]{Arnold1965}. More
specific to the Crab, both sources of magnetic free energy supplied by the
pulsar --- the striped wind and the large-scale magnetic hoops --- are seen to
dissipate by turbulent relaxation. Simulations reported in \cite{Zrake2016a}
indicate the alternating magnetic field pattern survives only of order its
proper Alfv{\'en} time before most of its energy has been dissipated into
turbulence, while simulations by \cite{Mizuno2011a} and also \cite{Mignone2013}
make analogous predictions regarding relaxation of the large-scale toroidal
magnetic field by kink instabilities. The general formalism we adopt for solving
the wind equations, and a precise explication of our dissipation model, is
presented in Section \ref{sec:equations}.

Much of the present paper focuses on validating this model for magnetic
dissipation in PWNe by matching features of the Crab Nebula. We explore the AC
and DC dissipation modes in turn, treating dissipation of striped wind in
Section \ref{sec:striped-wind}, and relaxation of the large-scale toroidal field
in Section \ref{sec:mean-field}. The purely AC case corresponds to a
hypothetical nebula powered by an orthogonally rotating pulsar, while the purely
DC case corresponds to an aligned rotator. While the Crab Pulsar's obliquity is
likely somewhere in between \citep[e.g.][]{Harding2008}, we show in Section
\ref{sec:mean-field} that the nebula's appearance can actually be described in
terms of a DC-only wind, in which relaxation of magnetic structures having the
scale of the termination shock yields a volume-average magnetization around
$10^{-2}$. By comparison, in a purely AC wind, dissipation runs too quickly and
the nebular magnetization falls to $\sim 10^{-9}$.

Our model is based on steady-state non-ideal MHD flow in spherical geometry.
Flow solutions are obtained using robust numerical methods for ordinary
differential equations. In this way we are able to simulate realistic wind
Lorentz factors and magnetization levels (each reaching $\sim 10^6$). Such
ultra-relativistic conditions cannot be accessed by typical Godunov-type MHD
codes. Our formalism also includes a self-consistent treatment of optically thin
synchrotron radiation, which we describe in Sections \ref{sec:conservation-laws}
and \ref{sec:synchrotron-radiation}. Energy and momentum losses to radiation
only marginally influence the dynamics of the Crab, as its radiative efficiency
is in the vicinity $20-30\%$ \citep{Hester2008}. However, we point out in
Section \ref{sec:gamma-ray-revshock} that radiative losses will become dominant
in certain scenarios of interest. In particular, forced magnetic reconnection in
a pulsar wind that is shocked at relatively close range due to confinement by a
dense medium is found to convert essentially $100\%$ of the wind power into
$\gamma$-rays. We comment on how this might impact theories of astrophysical
transients for which sustained energy injection by a millisecond magnetar is
invoked.

In Section \ref{sec:crab-flares} we discuss implications of our model for the
Crab Nebula $\gamma$-ray flares, and in Section \ref{sec:particle-spec} we
propose a means by which our model may yield improved calculations of the
non-thermal particle spectrum and evolution in PWNe.  Throughout the paper, we
utilize a notation in which $v$ denotes a speed (normalized by $c$) while $u =
\gamma v$ and $\gamma = 1 / \sqrt{1 - v^2}$ denote the corresponding four-speed
and Lorentz factor. Proper density of plasma is denoted by $\rho$, which is
implicitly multiplied by $c^2$ so it has dimensions of rest-mass energy per unit
volume. Variables with a subscript zero, e.g. $u_0$ indicate values at the base
of the wind.

\section{Equations of motion}
\label{sec:equations}

Here we develop equations for a stationary, relativistic plasma wind, subject to
radiative losses and turbulent magnetic reconnection. We adopt the toroidal MHD
approximation, in which the flow is radial and the magnetic field is
transverse. The flow is envisioned to contain a small-scale magnetic free energy
which is in a state of turbulent relaxation. The free energy, or ``eddy'' scale
is denoted by $\ell$, and unlike earlier analyses of magnetic thermalization in
relativistic outflows, we evolve that scale in a manner consistent with
numerical simulations of turbulent magnetic relaxation in relativistic systems
\citep{Zrake2014, Zrake2016}. This picture is adapted to the freely expanding
pulsar striped wind by initiating $\ell$ to the distance between magnetic
reversals ($\sim P_\star c$, where $P_\star$ is the pulsar rotation period), and
to the large-scale nebular magnetic field by initiating $\ell$ to the
termination shock radius. In the equations that follow, the flow variables are
understood to represent averages over scales that are larger than $\ell$, yet
smaller than the global system size.

\subsection{Conservation laws}
\label{sec:conservation-laws}

The equations of relativistic MHD are given by conservation of mass, energy,
momentum, and magnetic flux. Here, we will be using one-dimensional spherical
coordinates and assuming a steady-state flow, so that only derivatives with
respect to the radial coordinate $r$ are non-zero. Conservation of mass is given
in general by
\begin{equation*}
  \nabla_\mu \left( \rho u^\mu \right) = 0 \, ,
\end{equation*}
where $\rho$ is the comoving density, and $u^\mu$ is the four-velocity. In
spherically symmetric flow, the rate of mass loss per steradian, $f = r^2 \rho u
c$, is a constant at each radius, where $u = \gamma v$ is the radial
four-velocity. For optically thin plasma, conservation of energy and momentum is
given by
\begin{equation}
  \nabla_\mu T^{\mu \nu} = -u^\nu \dot \epsilon \, ,
  \label{eqn:rad-mhd}
\end{equation}
where $T^{\mu \nu}$ is the stress-energy tensor, and $\dot \epsilon$ is the
comoving emissivity, assumed to be isotropic in the plasma rest-frame. The time
component of Equation \ref{eqn:rad-mhd} is the energy conservation law, which
dictates that the change in wind luminosity $L = 4 \pi f \eta$ at each radius is
balanced by the radiative power. Here, $\eta = \gamma w$ is the luminosity per
unit mass loss, with $w$ denoting the total specific enthalpy. $\eta$ is
constant at each radius when radiation is neglected. The $r$-component of
Equation \ref{eqn:rad-mhd} is radial force balance,
\begin{equation*}
  \frac{d}{dr} \left[ r^2 \left( \rho w u ^2 + p + b^2/2 \right) \right] = -r^2
  u \dot \epsilon + 2 p r \, ,
\end{equation*}
which expresses cancellation of the total pressure gradient (ram, gas, and
magnetic) and inward magnetic tension force, and is the Bernoulli equation for
this system. Here, $b \equiv B_\phi / \sqrt{4 \pi} \gamma$ denotes the comoving
magnetic field (for notational convenience, $b$ is normalized so that magnetic
pressure is $b^2/2$). The term proportional to $\dot \epsilon$ represents wind
inertia carried away by photons.

Magnetic flux transport in a flow containing magnetic free energy is complicated
by non-ideal effects. Our approach here is to subsume the non-linear
reconnection physics into a dissipation function, denoted by $\dot \zeta$, which
prescribes the loss of magnetic flux over distance. The ideal MHD induction
equation,
\begin{equation}
  \partial_t \V B = \nabla \times (\V v \times \V B)
  \label{eqn:ideal-faraday}
\end{equation}
implies that $\frac{d}{dr}(r^2 b^2 u^2) = 0$, when the flow is toroidal and
stationary (in Equation \ref{eqn:ideal-faraday}, $\V v = c v
\hat{\V{r}}$). Dividing by $f$, we see that $\zeta = \sigma u$ is a
non-dissipative invariant. Here, $\sigma = b^2/\rho$ is the ratio of the wind's
electromagnetic to kinetic power. Note that this definition of $\sigma$ does not
include gas enthalpy in the denominator, and that the definition of
\cite{Kennel1984}, denoted as $\tilde \sigma \equiv \sigma / (1 + \mu)$ will be
used later on in Section \ref{sec:mean-field}.

To close the system we adopt a $\Gamma$-law equation of state, for which the gas
pressure $p = \rho e (\Gamma - 1)$, where $e$ is the thermal energy per unit
rest-mass energy. The corresponding specific entropy is $s = \ln
(p/\rho^\Gamma)$. Throughout we will use the value $\Gamma = 4/3$ appropriate
for relativistic gas particles. The total specific enthalpy is given by $w = 1 +
\mu + \sigma$, where $\mu = e + p/\rho = \Gamma e$ is the thermal enthalpy per
unit rest-mass energy.

The wind equations can now be written down as
\begin{eqnarray}
  df &=& 0 \label{eqn:df} \\
  \gamma \, d w + w \, d \gamma &=& -\frac{\gamma}{\rho} d\epsilon \label{eqn:deta} \\
  d \left( w u + \frac{\sigma}{2 u} \right) + \frac{d p}{\rho u} &=&
  -\frac{u}{\rho} d \epsilon \label{eqn:dp} \\
  u \, d\sigma + \sigma \, du &=& d\zeta \label{eqn:dzeta} \\
  \frac{d p}{p} - \Gamma \frac{d\rho}{\rho} &=& ds \, , \label{eqn:ds}
\end{eqnarray}
which are compact versions of the continuity equation, the energy equation, the
radial force balance, the phenomenological flux transport law, and the
thermodynamic entropy relation. All terms appearing on the right-hand side would
be zero when the wind is non-radiative and non-dissipative. The derivative $d$
may be with respect to $r$, or to the proper elapsed time of a fluid
element. Primes denote $d/dr$, while dots are comoving time derivatives,
e.g. $\dot \epsilon = u \epsilon'$.

Equations \ref{eqn:df} through \ref{eqn:ds} may be combined \footnote{Equation
  \ref{eqn:wind-ode} is obtained starting with Equation \ref{eqn:dp}. We then
  make the substitutions $\rho = f / r^2 u c$ and $dp = (dw - w_\rho d\rho -
  w_\sigma d\sigma) / w_p$, where the subscripts denote partial derivatives of
  the equation of state, written as $w(p, \rho, \sigma)$. We then insert
  expressions for $dw$ and $d\sigma$ obtained from Equations \ref{eqn:deta} and
  \ref{eqn:dzeta}. This leaves an expression in which the only remaining
  differentials are $du$, $dr$, $d\eta$, and $d\zeta$. We finally divide through
  by $dr$ and set $d\eta/dr = \dot \eta / u$ and $d\zeta/dr = \dot \zeta / u$.}
to yield the ordinary differential equation for the four-velocity,
\begin{equation}
  P \frac{d u}{d r} + Q \frac{u}{r} + R = 0 \label{eqn:wind-ode}
\end{equation}
where $P$, $Q$, and $R$ are given by
\begin{eqnarray*}
  P &=& \left(v_f^2 - v^2\right) \frac{\eta}{\Gamma - 1} \label{eqn:P}
  \\ Q &=& 2 \gamma \mu \label{eqn:Q} \\ R &=& \left(\frac{\Gamma - 2}{\Gamma -
    1} \right) \frac{1}{2 v} \, \dot \zeta - \dot \eta \, \label{eqn:R} ,
\end{eqnarray*}
and $v_f$ is the fast magnetosonic speed,
\begin{equation*}
  v_f^2 = \frac{\Gamma p + b^2}{\rho w} \, . \label{eqn:fast-speed}
\end{equation*}
Numerical wind solutions are obtained by integrating the unknowns $u$, $\zeta$,
and $\eta$ simultaneously in $r$, using Equation \ref{eqn:wind-ode}, and chosen
prescriptions for $\dot \eta$ and $\dot \zeta$.

The wind equations may also be arranged to give the rate of entropy
generation. By inserting the thermodynamic enthalpy relation $d\mu = e \, ds +
dp/\rho$ (which is equivalent to Equation \ref{eqn:ds}) into Equation
\ref{eqn:dp}, and then substituting Equations \ref{eqn:deta} and
\ref{eqn:dzeta}, we obtain
\begin{equation*}
  ds = \frac{1}{e} \left( \frac{d\eta}{\gamma} - \frac{d\zeta}{2 u} \right) \, .
\end{equation*}
This reflects that any non-ideal MHD processes $(d\zeta < 0)$ generate entropy
at the expense of otherwise frozen-in magnetic flux, while radiative losses
$(d\eta < 0)$ reduce entropy by cooling the plasma.

\subsection{General properties of the toroidal MHD wind}
\label{sec:general-properties}


It is worth pointing out certain mathematical features of Equation
\ref{eqn:wind-ode}. First, the toroidal MHD wind cannot go smoothly through a
fast magnetosonic point, regardless of how dissipation
occurs \footnote{Spherical MHD winds that accelerate smoothly through a fast
  point require that radial magnetic field and azimuthal velocity are non-zero
  \citep{Michel1969, Goldreich1970, Kennel1983}.}. Second, the post-shock flow
is always subsonic, and decelerates to a terminal speed proportional to the net
magnetic flux. We briefly explain these points here.

The lack of critical points can be shown by inspecting the polynomials $P$ and
$Q$. For $u$ to increase smoothly through $u_f$, the ratio $Q/P$ would need to
be finite there. But $P=0$ when $u=u_f$, so $Q$ would have to vanish
simultaneously. The condition $Q=0$ yields the quartic polynomial
\begin{equation}
  (\gamma^2 - 1) (\eta - \gamma)^2 - \gamma^2 \zeta^2 = 0 \, . \label{eqn:Q-polynomial}
\end{equation}
For Equation \ref{eqn:Q-polynomial} to have a root at $u_f$, the condition
\begin{equation}
  \eta^{2/3} - \zeta^{2/3} = 1 \label{eqn:degenerate-solution}
\end{equation}
must also be met. If Equation \ref{eqn:degenerate-solution} is satisfied, then
the wind has zero temperature and coasts along at the fast magnetosonic
speed. If Equation \ref{eqn:degenerate-solution} is not satisfied, then $P$ and
$Q$ have no simultaneous roots, and thus $u'$ cannot be finite where
$u=u_f$. These conclusions do not depend on the value of $R$, so however
dissipation or radiation may occur, the flow will not go smoothly through a fast
magnetosonic point.

Since $u'=0$ when $Q=0$ (assuming for the moment that $R=0$), the roots of
Equation \ref{eqn:Q-polynomial} represent asymptotic wind speeds. The subsonic
solution branch has a terminal speed given roughly by $\zeta / \eta$, so as
mentioned before, the post-shock flow moves away with a constant speed
proportional to the magnetic flux. Importantly, magnetic dissipation oppositely
affects the subsonic and supersonic solution branches. It is easily seen that $R
\ge 0$ (since $\dot \zeta \le 0$ and $1 < \Gamma < 2$), while $P$ is negative
for supersonic flow and positive for subsonic flow. Since $u' = -R/P$ (now
assuming that $Q=0$), it is clear that magnetic dissipation \emph{accelerates}
the supersonic (pre-shock) flow and \emph{decelerates} the subsonic (post-shock)
flow.


\subsection{Synchrotron radiation}
\label{sec:synchrotron-radiation}

Evolution of $\eta$ (the wind luminosity per particle) occurs only as the result
of radiative losses. If radiation were neglected, the plasma energy and momentum
would be conserved and the right-hand-side of Equation \ref{eqn:rad-mhd} would
be equal to zero. Here we assume that synchrotron is the dominant radiative
mechanism, that the nebula is optically thin to the emitted photons, that gas
pressure is isotropic, and that particles are mono-energetic, with the thermal
Lorentz factor given by $\gamma_{\rm th} = 1 + e$. The comoving synchrotron
luminosity per particle is given by \citep{Rybicki1979}
\begin{equation*}
  P_{\rm sync} = \frac{4}{3} \sigma_T c \, u_{\rm th}^2 u_B \, ,
\end{equation*}
where $\sigma_T$ is the Thomson cross section and $u_B \equiv \rho \sigma / 2$
is the magnetic energy density. The emissivity is given by $\dot \epsilon = n_e
P_{\rm sync}$, and recalling that $\dot \eta = -\gamma \dot \epsilon / \rho$
(Equation \ref{eqn:deta}), we have
\begin{equation}
  \dot \eta = - \frac{4}{9} \dot N \left( \frac{r_e}{r} \right)^2 \left(
  \frac{\gamma_{\rm th}^2 - 1}{v} \right) \sigma \, ,
\end{equation}
where $\dot N = 4 \pi f / m_e c^2$ is the pulsar's particle production rate,
$r_e$ is the classical electron radius, and $m_e = \rho / n_e c^2$ is the
electron mass. Further useful diagnostics to be encountered later on in Section
\ref{sec:mean-field-procedure} include the synchrotron frequency,
\begin{equation*}
  \nu_{\rm sync} = \frac{3}{4 \pi} \omega_g \gamma_{\rm th}^2 \, ,
\end{equation*}
where $\omega_g = v_{\rm th} / r_g$ and $r_g$ is the comoving electron
gyro-radius, and the dimensionless radiated power
\begin{equation*}
  \varepsilon_{\rm rad} = 1 - \eta / \eta_0 \, .
\end{equation*}

Our assumption that particles are distributed mono-energetically, $dn_e/d\gamma
= n_e \delta(\gamma - \gamma_{\rm th})$, may be dropped in a more sophisticated
analysis. For example, if one assumes power-law distributed particle energies,
$dn_e/d\gamma \propto \gamma^{-p}$ then an additional parameter $\gamma_{\rm
  max}$ needs to be specified independently. A reasonable choice for
$\gamma_{\rm max}$ is the energy of a particle whose gyro-radius is marginally
confined by the local turbulent eddies, $r_g \lesssim \ell$. Of course, that may
be an underestimate since particles experiencing Bohm diffusion $r_g \gg \ell$
continue to be accelerated by the second-order Fermi process. In modeling
emission from the Crab Nebula, \cite{Kennel1984a} assumed that particles reach
the energy at which they would be marginally confined within the termination
shock radius, $\sim \unit[\rm{few} \times 10^{17}]{cm}$.

\subsection{Turbulent magnetic relaxation}
\label{sec:turbulent-magnetic-relaxation}

Here we develop a prescription for the local rate $\dot \zeta$ of magnetic
dissipation in the pulsar wind nebula. Our starting assumption is that magnetic
free energy has a half-life $\tau = \ell / v_A$, where $\ell$ is the proper
scale of magnetic fluctuations (the ``eddy'' scale) and $v_A = (\sigma /
w)^{1/2}$ is the local Alfv{\'en} speed. Given $\tau$, the evolution of $\zeta$
is given simply by $\dot \zeta = -\zeta / \tau$. This expression is found by
first rearranging $\dot \zeta = \sigma \dot u + \dot \sigma u$, and then
separating $\dot \sigma$ into ideal and dissipative parts,
\begin{equation}
  \dot \sigma = -\frac{\sigma \dot u}{u} + \frac{\dot \zeta}{u} = \dot
  \sigma_{\rm ideal} + \dot \sigma_{\rm diss} \, .
\end{equation}
Equating the dissipative term $\dot \sigma_{\rm diss} \equiv \dot \zeta / u$
with $-\sigma/\tau$ yields the expression $\dot \zeta = -\zeta / \tau$.

So far, this procedure for treating dissipation is equivalent to that employed
by \cite{Drenkhahn2002} and by \cite{Giannios2007} for analysis of magnetic
dissipation in gamma-ray burst outflows. It is also similar to the formalism
developed by \cite{Lyubarsky2001} and \cite{Kirk2003} to characterize magnetic
reconnection in the pre-shock pulsar striped wind. The only significant
difference is that the latter authors utilized various microphysical
prescriptions to specify the speed $v_{\rm rec}$ of magnetic reconnection. In
Appendix \ref{app:reconnection-fronts} we show that our approach and theirs are
equivalent if $v_{\rm rec}$ is instead taken to be the Alfv{\'e}n speed, and the
comoving stripe separation is chosen as the eddy scale, $\ell \sim P_\star u c$.

A potentially significant effect, that was not accounted for in earlier studies,
is evolution of the eddy scale. In particular, growth of $\ell$ over time is now
understood to be a general feature of turbulent magnetic relaxation
\citep{Zrake2014, Brandenburg2015, Zrake2016, Campanelli2016}. Previously, the
``inverse energy transfer'' was thought to operate only when the field had a
significantly non-zero magnetic helicity measure \citep[see
  e.g.][]{Frisch1975}. MHD simulations presented in \cite{Zrake2014} demonstrate
that a magnetic field, that is initially tangled isotropically at a small scale,
relaxes according to $\dot \sigma \sim -\sigma / \tau$, while $\dot \ell \sim
\ell / \tau$. Turbulent motions are sustained at roughly the comoving Alfv{\'e}n
speed. Over time, the cascade slows because the Alfv{\'en} speed drops, and also
because the eddy scale grows. This process yields self-similar temporal and
spectral evolution that persists while $\ell$ remains smaller than a global
system scale (the nebula in the present context).

The scaling behavior of turbulent relaxation reported in \cite{Zrake2014} was
limited to conditions where the plasma was overall at rest in a periodic
box. However, we are interested here in decaying turbulence that is embedded
into a flow, and whose scale is thus influenced by the expansion or compression
of that flow as it accelerates or decelerates. In other words, evolution of
$\ell$ is driven by adiabatic effects, and simultaneously by the dissipative
action of the cascade. For this reason, we choose to characterize eddies based
on their mass $m$, which remains fixed under the adiabatic effects alone. The
eddy mass and scale are related through the ambient density $\rho$, and the
assumption that eddies are isotropic; $m \equiv \rho \ell^3$. When the flow is
not expanding or contracting, the two laws $\dot m \sim m / \tau$ and $\dot \ell
\sim \ell / \tau$ say the same thing (with appropriate prefactors). Evolution of
$m$ in this way simply states that magnetic structures intend to reduce their
energy by merging with one another. Merging between eddies proceeds over a
dynamical time, and results in an eddy with double the mass. The process repeats
itself hierarchically, while $\tau$ adjusts to the local values of $\ell$ and
$v_A$,
\begin{eqnarray}
  \dot \zeta &=& -\zeta / \tau \label{eqn:turbulent-relaxation} \\
  \dot m &=& m / \tau \nonumber \\
  \tau &=& \ell / v_A \nonumber \\
  \ell &=& (m / \rho)^{1/3} \nonumber \, .
\end{eqnarray}
More details on the turbulent magnetic relaxation picture, including a sensible
explanation of the inverse energy transfer, are provided in Appendix
\ref{app:plasmoid-cascade}.



\section{Turbulent dissipation of the striped wind}
\label{sec:striped-wind}

\subsection{Turbulent dissipation in the free-expanding wind}
\label{sec:pre-shock-stripe-dissipation}

\begin{figure}
  \centering
  \includegraphics[width=3.55in]{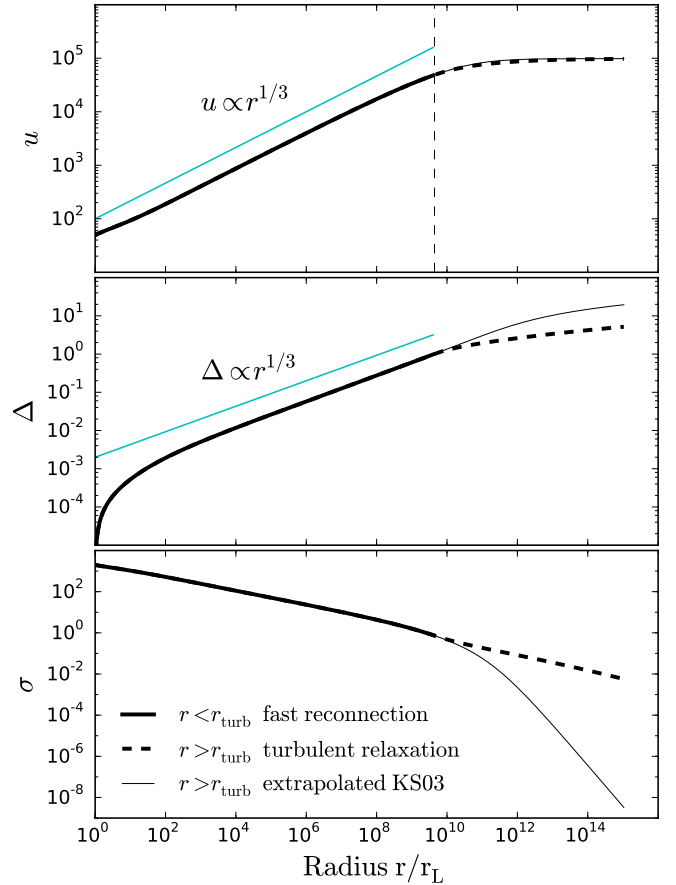}
  \caption{Shown here is the four-velocity $u$ (top), and magnetization $\sigma$
    (bottom) as a function of distance, for different models of a dissipative
    MHD wind with $\sigma_0 = 2000$. The thick solid curve is the
    ``fast-reconnection'' picture described in \cite{Kirk2003}. The vertical
    dashed line indicates the radius $r_{\rm turb}$ at which a dynamical time
    has elapsed in the comoving frame, and where the fast reconnection picture
    should break down. The dashed line shows evolution according to the
    turbulent relaxation picture.}
  \label{fig:FastVsTurbulentReconnection}
\end{figure}

Winds driven by obliquely rotating pulsars contain ``stripes'' around the
equatorial region. The stripes are reversals of the azimuthal magnetic field,
separated by a half-wavelength $\lambda = P_\star v / 2$ \citep{Michel1971,
  Michel1982}. Magnetic reconnection across the reversals has been considered as
a possible mechanism by which the wind's electromagnetic energy may be
dissipated upstream of the termination shock \citep{Coroniti1990,
  Michel1994}. However, those treatments neglected feedback of the dissipated
magnetic energy on the global flow. Energy conserving analyses
\citep{Lyubarsky2001, Kirk2003} revealed that dissipated magnetic energy goes
toward accelerating the flow, due to reduction of inward-pointing magnetic hoop
stress. As the flow moves faster, relativistic time dilation suppresses the
dissipation rate as seen in the pulsar frame. The Crab Pulsar's wind becomes
weakly magnetized by the time it reaches the termination shock if it is at the
upper end of believable mass-loading $\dot N \gtrsim \unit[10^{40}]{s^{-1}}$.

The analysis given in \cite{Lyubarsky2001} was based on a plane-parallel
reconnection model, in which cold, magnetized plasma fills the volume between
hot, unmagnetized current layers. Hot current layers expand into the cold plasma
at a speed $v_{\rm rec}$, consuming magnetic energy as they proceed. The
fraction of volume occupied by hot plasma, denoted as $\Delta$, indicates how
much of the initial magnetic energy has been consumed, $\zeta = (1 - \Delta)
\zeta_0$. To first order in both the expansion speed $v_{\rm rec}$ and the
inverse Lorentz factor $\gamma^{-1}$, $\Delta$ evolves according to (see
Appendix \ref{app:reconnection-fronts})
\begin{equation}
  \dot \Delta = 2 \frac{v_{\rm rec}}{\lambda \gamma} + \mathcal{O}(\gamma^{-2}) +
  \mathcal{O}(v_{\rm rec}^2) \, . \label{eqn:reconnection-front-speed}
\end{equation}
Thus, Lyubarsky's prescription and ours are made equivalent by the
identifications $\ell \rightarrow \lambda \gamma$, $v_{\rm rec} \rightarrow v_A / 2$,
and $\dot \Delta \rightarrow \tau^{-1}$, and by replacing $\zeta = (1 - \Delta)
\zeta_0$ with $\dot \zeta = -\dot \Delta \zeta$, so that $\Delta$ is formally
allowed to exceed unity. Conveniently, $\Delta$ may here be interpreted as the
number of elapsed Alfv{\'e}n times felt by a fluid parcel.

It was argued in \cite{Zrake2016a} that the planar geometry of the current
layers would only persist until order one comoving Alfv{\'e}n times had elapsed,
$\Delta \sim 1$. Subsequently the striped magnetic field becomes isotropic and
dissipation should be described by turbulent relaxation. Thus, the onset of
isotropic turbulence accompanies the transition to particle-dominated flow,
which again, occurs upstream of the shock only if the wind is well
mass-loaded. Nevertheless, we will explore briefly evolution of the striped
wind, if it were allowed to expand freely beyond the turbulence transition. We
do this by setting $\ell$ to $\lambda \gamma$ until $\Delta = 1$, and henceforth
evolving $\ell$ according to the turbulent relaxation model. For as long as
$\Delta \ll 1$, $\sigma$ remains large and thus $v_A \approx c$. We expect the
asymptotic solutions to obey the same scaling as the ``fast reconnection''
picture presented in \cite{Kirk2003}, because that also adopts constant
reconnection speed (that of sound $c / \sqrt{3}$ in the hot plasma).

Figure \ref{fig:FastVsTurbulentReconnection} illustrates this equivalence for a
hypothetical wind having a magnetization $\sigma_0 = 2000$, and which continues
to $10^{15} r_L$ without passing through a termination shock. We see that indeed
$\Delta$ and $u$ both evolve $\propto r^{1/3}$. Beyond $\Delta = 1$, the
solution is integrated in two ways: (1) by extending the fast reconnection
picture, where $\ell$ remains equal to $\lambda \gamma$ and (2) by using the
turbulent relaxation picture where $\ell$ increases according to Equation
\ref{eqn:turbulent-relaxation}. Since roughly half of the magnetic energy has
already been spent by the time $\Delta \sim 1$, further evolution of $u$ is not
affected much by how dissipation is prescribed where $\Delta > 1$. Magnetization
$\sigma$, on the other hand, evolves much more slowly when turbulent growth of
$\ell$ is accounted for.

\subsection{Forced reconnection at the shock}
\label{sec:post-shock-stripe-dissipation}

\begin{figure}
  \centering \includegraphics[width=3.55in]{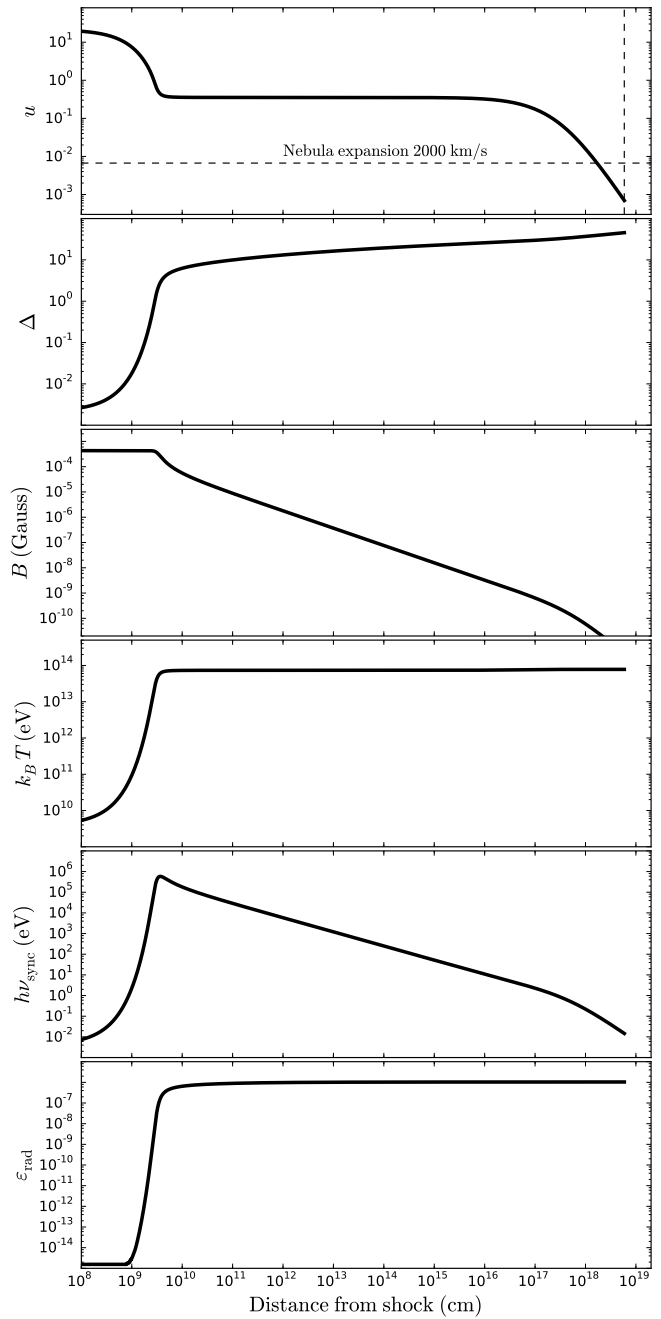}
  \caption{Evolution of the post-shock pulsar striped wind. Shown here, from top
    to bottom, is the radial four-velocity $u$, cascade number $\Delta$,
    magnetic field strength $B$, electron temperature $k_B T$, photon energy $h
    \nu_{\rm sync}$, and radiated power $\varepsilon_{\rm rad}$, all as a
    function of distance downstream of the shock, for particular wind parameters
    $L_\star = \unit[5 \times 10^{38}]{erg/s}$, $\dot N =
    \unit[10^{36}]{s^{-1}}$, and $\sigma_0 = 2 \times 10^5$. Dissipation occurs
    in the ``fast reconnection'' regime upstream where $\Delta < 1$, and in the
    turbulent regime once $\Delta \ge 1$. The horizontal dashed line in the
    top-most panel indicates the nominal nebula expansion speed $u_{\rm neb} =
    \unit[2000]{km/s}$, while the vertical dashed line marks the nominal nebula
    boundary $r_{\rm neb} = \unit[2]{pc}$.}
  \label{fig:PostshockStripedWind}
\end{figure}



Now we consider dissipation of the striped wind beyond the termination shock,
assuming that it arrives there well magnetized. At the shock, the flow is
compressed and decelerated below the fast magnetosonic speed. So, as discussed
in Section \ref{sec:general-properties}, any dissipation that occurs beyond it
only further decelerates the flow. Such ``forced reconnection'' has been studied
in depth for its possible role in energizing at least some of the Crab Nebula's
radio-emitting, non-thermal electrons. However, as pointed out by
\cite{Lyubarsky2003}, the post-shock temperature is likely high enough that
electron gyro-radii exceed the stripe wavelength, $r_g > \ell$. Electron heating
in this regime has been studied using kinetic simulations by
\cite{Lyubarsky2008}. Forced reconnection of the striped wind has also been
examined in a regime where post-shock kinetic scales remain smaller than the
stripe separation, in both kinetic \citep{Sironi2011} and hydromagnetic
\citep{Takamoto2012} settings. Here, we analyze the post-shock forced
reconnection using our model, even though turbulent dissipation is probably not
an appropriate treatment when the eddy scale is microscopic. Still, this
approach extends the analysis of \cite{Lyubarsky2003}, by resolving the forced
reconnection zone. This approach also extends the analyses of
\cite{Lyubarsky2008, Sironi2011, Takamoto2012}, which do resolve the post-shock
reconnection zone, but do not account for the flow's spherical geometry or its
macroscopic evolution throughout the nebula. Our aim is to determine the
magnetization and radiative efficiency of the nebular flow, at latitudes where
stripes may be fully dissipated. This applies to the equatorial belt in the Crab
Nebula, or at all latitudes in a hypothetical nebula powered by an orthogonally
rotating pulsar.

We choose a set of wind parameters, $L_\star = \unit[5 \times 10^{38}]{erg/s}$,
$\dot N = \unit[10^{36}]{s^{-1}}$, and $\sigma_0 = 2 \times 10^5$, intended to
illustrate extreme electron heating and photon production in the forced
reconnection. Such small mass-loading implies that the each particle must bear a
relatively greater portion of the wind power, such that the post-shock
temperature and thus synchrotron frequency is higher. The freely expanding wind
is assumed to dissipate in the fast reconnection regime described in Section
\ref{sec:pre-shock-stripe-dissipation}, because for these parameters, $\Delta$
remains small ($\lesssim 10^{-2}$) out to the shock. At the distance $r_{\rm ts}
= \unit[3 \times 10^{17}]{cm}$, we solve the Kennel-Coroniti jump conditions
(see Equation \ref{eqn:kc-jump-conditions}) and continue integrating the
solution on the downstream side of the shock. There, $\Delta'$ increases
dramatically such that $\Delta$ goes through unity after a few wavelengths $\sim
\unit[10^9-10^{10}]{cm}$. Once $\Delta > 1$ we switch to the turbulent
relaxation model.

A number of points are illustrated by the solution, shown in Figure
\ref{fig:PostshockStripedWind}. First, dissipation (and thus deceleration) occur
rapidly on the downstream side of the shock, even though turbulent relaxation is
slowed by the growth of $\ell$. Since we have assumed zero residual
(stripe-averaged) magnetic field, reconnection proceeds toward arbitrarily small
values of the magnetic field. This has two immediate consequences --- (1) that
the flow decelerates well below the nebula expansion velocity $u_{\rm neb}
\approx \unit[2000]{km/s}$, such that the outer boundary condition cannot be
satisfied, and (2) that the magnetic field at large distance lies in the range
of $\unit[10^{-9}]{G}$, much weaker than what is thought to exist in the Crab
Nebula, $\unit[10^{-4}-10^{-3}]{G}$. Thus, orthogonal rotation of the Crab
Pulsar is incompatible with our dissipation model. If a residual magnetic field
is included \footnote{A residual magnetic field $\bar \zeta$ may been
  implemented by replacing Equation \ref{eqn:turbulent-relaxation} with $\dot
  \zeta = -(\zeta - \bar \zeta) / \tau$. Such solutions have been examined but
  are not shown here, because they are identical, in the context of the striped
  wind, to those obtained with the dissipative shock jump condition of
  \cite{Lyubarsky2003}.}, then only that part remains after a short distance
beyond the shock. A residual field might well be interpreted as the
magnetization invoked by \cite{Kennel1984} to match the nebula
expansion. However, any such residual field at a given latitude should also be
prone to reconnection by mixing across the nebula equator or succumbing to kink
instabilities at high latitude. We analyze relaxation of the large-scale
residual magnetic field in Section \ref{sec:mean-field}.


Figure \ref{fig:PostshockStripedWind} illustrates a number of points related to
synchrotron radiation from the pulsar wind. First, effectively no radiation
comes from the freely expanding wind; the synchrotron cooling time for streaming
elections is vastly longer than their adiabatic cooling time. This point is
already well appreciated, and known empirically from the Crab's
``underluminous'' zone \cite[e.g.][]{Hester2002}. Second, electrons are only
heated up to $\sim \unit[10^{10}]{eV}$ by the shock itself, but may be further
heated (at least for the low mass-loading chosen here) up to $\sim
\unit[0.1]{PeV}$ in the forced reconnection immediately downstream. The
associated photon energies reach to the MeV range. However, the magnetic field
dissipates at short range to much lower values, and only $\sim 10^{-6}$ of the
pulsar luminosity is ultimately converted into $\gamma$-rays.

The wind parameters chosen for this example are intentionally chosen to
represent an extreme case of electron heating in the forced reconnection. For
more realistic parameters, say $\dot N \sim 10^{38}$, the post-shock electron
temperatures are typically in the TeV range, with associated photon energies in
the UV to soft X-ray. However, the choice of wind parameters has little bearing
on the profile of flow velocity and magnetic field strength at distances larger
than $\sim \unit[10^{17}]{cm}$. In particular, we have not identified any
relevant wind parameters which yield a realistic radiative efficiency, again
because the short-wavelength oscillating magnetic field dissipates too quickly;
producing the high radiative efficiency inferred for the Crab requires a
residual, or at least slowly dissipating component of the magnetic field. In the
next section, we show that our turbulent relaxation model can match the outer
boundary condition and radiative efficiency if applied to the large-scale
toroidal magnetic field, rather than the striped wind.



\section{Turbulent relaxation of the large-scale field in the Crab Nebula}
\label{sec:mean-field}

We now turn to the question of how the large-scale toroidal magnetic field
relaxes and dissipates in the volume of the nebula. This analysis intends to
characterize the post-shock flow at moderate to high latitudes, where the
oscillating magnetic field is small or zero. We thus evolve the freely expanding
flow without any dissipation, while the post-shock flow is evolved using the
turbulent relaxation prescription. We fix the wind luminosity $L_\star = \unit[5
  \times 10^{38}]{erg/s}$, which now stands for power supplied to the nebula in
the form of non-oscillating magnetic field, and should thus, strictly speaking,
be smaller by a factor of $\sim 3$ \citep[][assuming $45^\circ$
  obliquity]{Komissarov2012} than the pulsar spin-down power. The radiative
efficiency is set to a nominal value of $\varepsilon_{\rm neb} = 0.4$, which is
marginally higher than current estimates in the range of $20-30\%$
\citep{Hester2008}. The eddy scale near the inner edge of the nebula, $\ell_{\rm
  sh}$ is left as a free parameter, to be determined by appropriate matching of
the outer boundary condition. The corresponding rate $\tau_{\rm sh}^{-1}$ may be
interpreted loosely as the growth rate of kink instabilities operating on the
toroidal field around nebula's X-ray core \citep{Begelman1998}. At larger
distances the dissipation time-scale will increase. If the model is realistic,
the volume average $\bar \tau$ will be close to $\sim \unit[80]{yr}$, estimated
by \cite{Komissarov2012} from calorimetric considerations.

\subsection{Numerical procedure}
\label{sec:mean-field-procedure}

We identify a family of wind solutions, one for each value of $\ell_{\rm sh}$,
that matches the nebula expansion speed $u_{\rm neb} = \unit[2000]{km/s}$ and
radiative efficiency $\varepsilon_{\rm neb} = 0.4$. We do this by numerically
integrating 144 wind solutions, tabulated on an evenly spaced $12 \times 12$
grid in $x \equiv \log \dot N$ and $y \equiv \log \sigma$. One such grid is
generated for each of 30 different values of $\ell_{\rm sh}$. On each grid,
[$u_{\rm neb}$, $\varepsilon_{\rm neb}$] is recorded at all lattice points
$(x,y)$.  We then prolong the grid by a factor of 8 in both variables using
third-order splines. Next we construct a set of level surfaces
$[x_{u,\varepsilon}(s)$, $y_{u,\varepsilon}(s)]$, parameterized in arc-length
$s$, along which $u_{\rm neb}$, $\varepsilon_{\rm neb}$ take on constant
values. Finally, we determine the intersection between the level surfaces by
minimizing the distance function $\sqrt{[x_u(s_u) -
    x_\varepsilon(s_\varepsilon)]^2 + [y_u(s_u) -
    y_\varepsilon(s_\varepsilon)]^2}$ over ($s_u$, $s_\varepsilon$). Figure
\ref{fig:ExpRadContourPlot} shows an example of one of the solution grids.

\subsection{Results}
\label{sec:mean-field-results}

\begin{figure}
  \centering
  \includegraphics[width=3.55in]{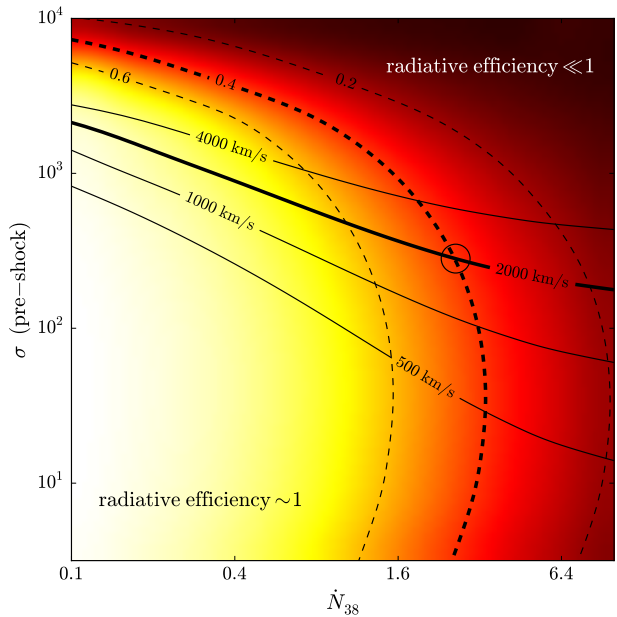}
  \caption{Relief plot showing the radiative efficiency $\varepsilon_{\rm neb}$
    of the post-shock flow in the Crab Nebula, as a function of the wind
    magnetization $\sigma$ and particle production rate $\dot N = \unit[10^{38}
      \dot N_{38}]{s^{-1}}$. The color indicates the fraction of pulsar
    spin-down power that has been converted into synchrotron photons once the
    flow reaches the nebula boundary at $\unit[\sim 2]{pc}$. Dashed contours are
    the level surfaces of radiative efficiency $\varepsilon_{\rm neb}$. The
    solid contours represent level surfaces of $u_{\rm neb}$, the nebula
    expansion speed at its boundary. The open circle at the intersection of the
    two heavier contours indicates the values of $\sigma$ and $\dot N$, that are
    favored by the turbulent relaxation model, given the fiducial values
    $\varepsilon_{\rm neb} = 0.4$ and $u_{\rm neb} = \unit[2000]{km/s}$.}
  \label{fig:ExpRadContourPlot}
\end{figure}

\begin{figure}
  \centering
  \includegraphics[width=3.55in]{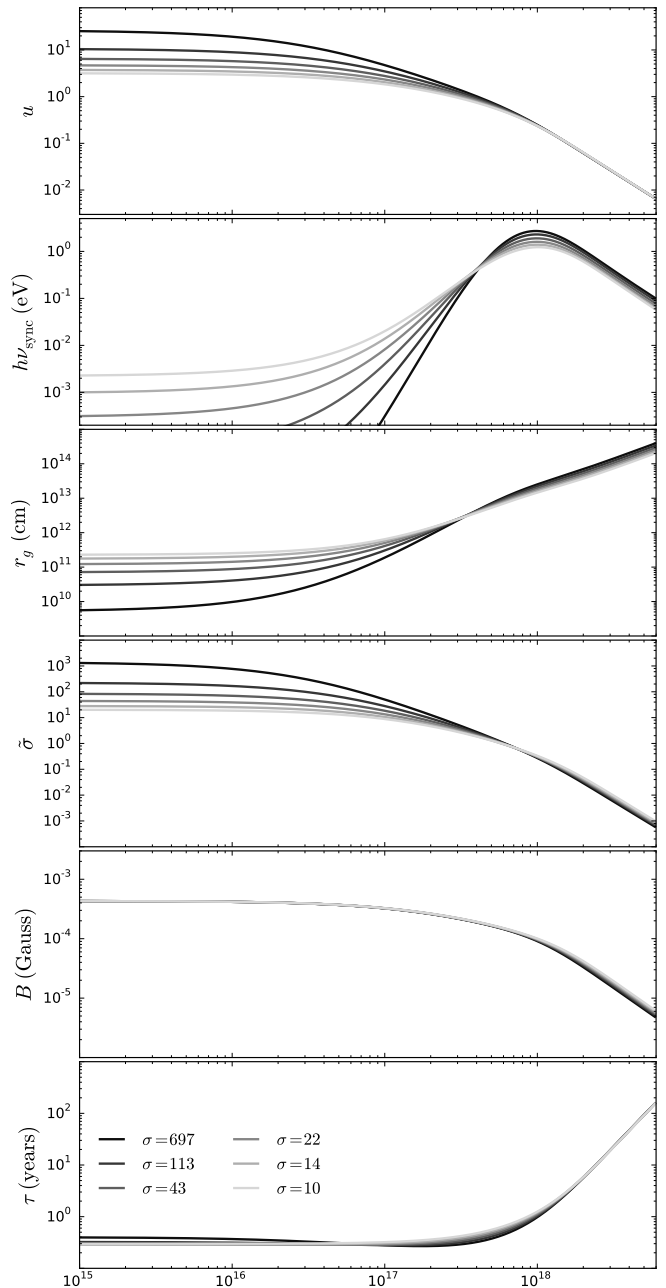}
  \caption{Solutions of the post-shock nebular flow, when dissipation occurs due
    to large-scale turbulent processes. Shown are six solutions with upstream
    magnetizations $\sigma_0$ between 10 and 697, but that all satisfy the
    nebula boundary condition $u_{\rm neb} = \unit[2000]{km/s}$ and have a
    radiative efficiency $\varepsilon_{\rm neb} = 0.4$. Shown, from top to
    bottom, is the four velocity $u$, photon energy $h \nu_{\rm sync}$, electron
    gyro-radius $r_g$, magnetization $\tilde \sigma \equiv \sigma / (1 + \mu)$,
    magnetic field strength $B$, and local dissipation time-scale $\tau$, all as
    a function of distance downstream of the shock.}
  \label{fig:SolutionPanels}
\end{figure}

Exploration of the parameter space reveals that the outer boundary condition and
the radiative efficiency can only be matched simultaneously when $\ell_{\rm sh}$
is quite near (within $\sim 30\%$ of) $r_{\rm ts} = \unit[3 \times
  10^{17}]{cm}$, and when $\dot N = \unit[10^{38}]{s^{-1}}$ to within a factor
of $\sim 5$. On the other hand, the pre-shock value of $\sigma$ is found to be
rather sensitive to $\ell_{\rm sh}$, ranging from $10$ when $\ell_{\rm sh} =
\unit[2.8 \times 10^{17}]{cm}$ to $697$ when $\ell_{\rm sh} = \unit[3.9 \times
  10^{17}]{cm}$. Figure \ref{fig:ExpRadContourPlot} shows an example solution
grid for which $\ell_{\rm sh} = \unit[3.5 \times 10^{17}]{cm}$. The level
surfaces of $u_{\rm neb}$ and $\varepsilon_{\rm neb}$ intersect at $\dot N =
\unit[2.6 \times 10^{38}]{s^{-1}}$ and $\sigma_0 = 285$. Radiative efficiency
increases with lower mass-loading because each particle must bear a greater
portion of the wind's dissipated magnetic energy, and thus attains a higher
thermal Lorentz factor. The nebula magnetic field strength is found to be
relatively insensitive to $\dot N$.

Figure \ref{fig:SolutionPanels} shows six of the matching solutions for
different values $\ell_{\rm sh}$, labeled for the value of $\sigma_0$ that
satisfies the boundary conditions. The different solutions are largely
distinguished by the post-shock speed, which for the higher magnetization value
remains at $u \sim 10-20$ over $\sim \unit[10^{16}-10^{17}]{cm}$ beyond the
shock. For all solutions, the electron gyro-radius $r_g$ (and indeed the skin
depth, not shown) is larger than the oscillating magnetic field wavelength,
suggesting that any stripes would dissipate without energizing non-thermal
electrons \citep{Sironi2011}. However, the kinetic scales remain everywhere much
smaller than the eddy scale, indicating that MHD turbulent relaxation is an
appropriate description of the mean field dissipation. Furthermore, non-thermal
particle acceleration may be successful in the turbulent reconnection, since
there the magnetic fluctuation scale is macroscopic.

Solutions are indistinguishable in the profile of magnetic field strength, which
remains between 100 and $\unit[400]{\mu G}$ out to $\unit[10^{18}]{cm}$ and then
declines to $\unit[10]{\mu G}$ by the nebula's outer edge. The volume average of
the nebula magnetic field is very nearly $\unit[60]{\mu G}$ in all our models,
about half that required by the one-zone model of \cite{Meyer2010}. The photon
energy peaks at $\unit[10^{18}]{cm}$ in all models, and varies from
$\unit[1-3]{eV}$ as $\sigma_0$ increases (and $\dot N$ decreases). The Crab
spectrum peaks at around $\unit[4]{eV}$, so mass loadings at the lower end of
our parameter range, $\dot N_{38} = 2.5$, are weakly favored. Electron
temperatures are from $\unit[100-200]{GeV}$ around $r = \unit[10^{18}]{cm}$, and
the gyro-radii are from $\unit[10^{13} - 10^{14}]{cm}$. Meanwhile, the turbulent
eddy scale (not shown in the figure) increases from $\sim r_{\rm ts}$ by a
factor of roughly 10, to $\sim r_{\rm neb}$ at the outer edge, so the relaxation
picture becomes marginally inapplicable there. The dissipation time-scale varies
from about 4 months near the nebula core, to over 100 years at the outer
edge. The volume average is $\bar \tau \approx 52$ years for all models,
slightly lower than the 80-year dissipation time scale determined in
\cite{Komissarov2012}.


\section{Discussion and conclusions}
\label{sec:discussion}

We have developed a model for turbulent magnetic reconnection in a pulsar wind
nebula. The model is based on stationary, one-dimensional MHD flow which embeds
a freely relaxing, small-scale magnetic field. Such characterization of the
internal magnetic field allows for a dissipation model that is supported by
recent developments in the theory of relativistic magnetic reconnection and
turbulent relaxation. We examined separately dissipation of the striped wind in
both the freely expanding and post-shock flow, as well as turbulent relaxation
of the mean magnetic field beyond the termination shock. Applied to the Crab
Nebula, this model reproduces the expansion speed, radiative efficiency, and
peak photon frequency within a one-parameter family of solutions, and without
invoking unrealistically low values of the upstream magnetization, as required
in the historical non-dissipative models. Our model strongly favors a pulsar
particle production rate $\dot N = \unit[2.5 \times 10^{38}]{s^{-1}}$, and a
magnetic thermalization time scale which evolves from roughly 4 months (the
light-travel time of the wind termination shock radius) near the nebula core, to
100 years near its outer edge. Such slowing of the turbulent cascade rate with
distance is not put in by hand, but results from scaling laws derived from
three-dimensional numerical simulations of turbulent magnetic relaxation
\citep{Zrake2014}. We have also introduced a simple formalism for incorporating
optically thin synchrotron losses into steady-state MHD winds. Though not
critical in the dynamics of the Crab Pulsar wind, we will point out in Section
\ref{sec:gamma-ray-revshock} that the radiative losses may dominate the reverse
shock dynamics in the striped wind of a ``millisecond magnetar''.

\subsection{Limitations}

The model we have developed here is overly simplistic in a number of ways.
First, we do not account for any non-spherical geometry. In this approximation,
the equations of motion and the shock jump conditions are formally valid in
transverse MHD only at the equator. More realistic, two-dimensional shock jump
conditions have been analyzed in connection with the Crab's inner knot feature
\citep{Yuan2015, Lyutikov2016}. Also, in treating the radiative losses we
assumed that particles were mono-energetic, whereas \cite{Kennel1984a} produced
integrated spectra by assuming that particles occupy a power-law in energy. In
principle, we could have done the same here, but that would require a particle
spectral index to be selected by hand. Given that the magnetic field strength
and coherence scale are both predicted by the model, the particle spectrum could
be truncated at the energy where gyro-radii would exceed the local eddy scale. A
more sophisticated treatment along these lines may be pursued in a future work.

\subsection{Crab $\gamma$-ray flares}
\label{sec:crab-flares}

The Crab's $\gamma$-ray flares \citep{Tavani2011, Abdo2011} have been widely
interpreted as the signature of exceptionally powerful reconnection episodes
\citep{Uzdensky2011, Clausen-Brown2012, Cerutti2012, Cerutti2014, Cerutti2014a,
  Nalewajko2016, Yuan2016, Lyutikov2016a}. Reconnection is most promising to
explain flares if it occurs where the plasma is strongly magnetized. As pointed
out by \cite{Lyubarsky2012}, such a region is expected to occur in the $z$-pinch
at high latitudes. In his model, the core of the pinch would be of the order
$\unit[10^{16}]{cm}$, roughly the scale of the emitting region as implied by the
$\sim \unit[10]{hr}$ flare duration. Despite the rather artificial geometry of
our own model, it also predicts a strongly magnetized region region, $\tilde
\sigma \lesssim 10^3$ of scale $\unit[10^{16}-10^{17}]{cm}$, when the upstream
magnetization is high. Interestingly, in that region (just past the termination
shock, see Figure \ref{fig:SolutionPanels}) plasma moves radially outward with a
Lorentz factor $\lesssim 20$. The implied Doppler beaming could help account for
the spectral cutoff, which for some flares exceeds the radiation reaction limit
$\sim \unit[100]{MeV}$ \citep[e.g.][]{Buehler2012}. In our model, the post-shock
flow is relativistic and strongly magnetized only where stripes are
negligible. Where they dominate, the magnetization decreases and the flow
decelerates immediately beyond the shock in the forced reconnection (see Figure
\ref{fig:PostshockStripedWind}). This may still be consistent with a scenario
discussed in \cite{Arons2002}, where the flaring region lies marginally upstream
(or perhaps within the kinetic structure) of the termination shock, where the
radial Lorentz factor remains high.

\subsection{Evolution of the PWN particle spectrum}
\label{sec:particle-spec}

We hope that the model developed here will find utility in multi-zone modeling
of particle transport and acceleration in PWNe. This could be useful in modeling
of PWNe for which multi-band, spatially resolved observations are available,
such as MSH 15-52 \citep[e.g.][]{An2014} and 3C 58, not to mention the Crab. It
could also help determine the rate of positron leakage from the nebula, and help
assess the importance of PWNe as galactic positron sources. Unlike one-zone
models \citep{Gelfand2009, Bucciantini2011} for which the spectral energy
distribution of injected electrons is a model parameter, a multi-zone treatment
could possibly predict the spectrum from humbler assumptions. For example, one
might solve an advection-diffusion equation, where the advective part is given
by the flow velocity $u$, and the spatial diffusion by the eddy scale $\ell$ and
fluctuating velocity $v_A$. The latter yields a turbulent diffusivity, while the
local specific heating rate $\dot \zeta$ may be used to normalize energy gains
by the electron population. This may be taken up in a future work.

\subsection{EM counterparts from ``millisecond magnetars''}
\label{sec:gamma-ray-revshock}

Rapidly rotating, strongly magnetized neutron stars have been invoked in
connection with a variety of observed and hypothetical astrophysical
transients. Many $\gamma$-ray bursts, both long and short, exhibit an extended
X-ray emission plateau, which has been widely interpreted as the signature of
sustained energy injection by a nascent PWN. Generally, one imagines the
nebula to be confined by a massive shell, either of stellar ejecta in the case
of long bursts and superluminous supernovae, or material dynamically expelled
from a binary neutron star merger in the case of short bursts. In either
scenario, the resulting emission is expected to be more isotropic than the
$\gamma$-ray burst itself, and is thus interesting as a possible electromagnetic
counterpart to gravitational wave detections of binary neutron star coalescence.

Conditions in these hypothetical PWNe are qualitatively distinct in a couple of
ways from those of the Crab Nebula. The radiative energy density
$\varepsilon_{\rm rad}$ in the nascent nebula could be so high as to initiate a
cascade of $e^\pm$ pair production, so that mass conservation $df=0$ does not
apply. As a result, the plasma density may be high enough to trap synchrotron
photons, so that $d \eta = 0$ even when $P_{\rm sync} \ne 0$. Pair production is
robust when the nebula compactness parameter
\begin{equation}
  \ell_c = (\varepsilon_{\rm rad} / m_e c^2) \, \sigma_T \, \delta R_{\rm neb}
  \label{eqn:cp}
\end{equation}
is high. As we saw in Section \ref{sec:post-shock-stripe-dissipation} the
pre-shock flow is cold and non-radiative, so pair production should only
commence in the region of width $\delta R_{\rm neb}$ between the reverse shock
and the confining shell. Equation \ref{eqn:cp} indicates that if some photons
were to leak out from the nebula, then pair production would be suppressed for
two reasons. The first, of course, is that leakage reduces $\varepsilon_{\rm
  rad}$ and thus $\ell_c$ directly. The second is that loss of radiation
pressure in the shocked plasma allows the reverse shock to advance toward the
shell, reducing $\delta R_{\rm neb}$. This further decreases the nebula optical
depth, meaning that a slightly ``deflated'' nebula only becomes more leaky.

Of course, the details of radiation leakage from nascent PWNe are beyond the
present scope. Nevertheless, we make two observations here that could motivate
future more detailed calculations. First, we point out that under conditions
relevant to binary neutron star mergers (those yielding a millisecond magnetar),
the synchrotron efficiency of the reverse shock can go up to $100\%$, with all
the radiation produced above $\unit[100]{MeV}$. Photon losses may become
relevant at such high energies because Klein-Nishina effects reduce the optical
depth of the shocked plasma and confining shell. Also, the relevant shell albedo
would be that of $\gamma$-rays, while X-ray albedo has been adopted in earlier
analyses \citep{Metzger2013a, Metzger2014, Kasen2016}. Second, we argue that the
pair cascade might in some cases appear intermittently or not at all.

Consider the special case of a stable millisecond magnetar formed in a binary
neutron star merger. We adopt nominal source parameters $L = \unit[5 \times
  10^{47}]{erg/s}$, $\dot N = \unit[10^{48}]{s^{-1}}$, and $\sigma_0 = 5 \times
10^3$, and place the ejecta at $R_{\rm ej} = 0.1 c \times \unit[1]{hr} \approx
\unit[10^{13}]{cm}$, moving outwards at $0.1c \approx \unit[30,000]{km/s}$. Now
suppose the pulsar is an orthogonal rotator, so that its electromagnetic power
is all in the striped wind as we analyzed in Section
\ref{sec:striped-wind}. With those parameters fixed, vary the shock location
until the flow matches smoothly onto the ejecta. If the plasma is optically
thick, then synchrotron losses should be ignored as photons are trapped and
contribute to the gas pressure with the same equation of state, $\Gamma=4/3$ as
the plasma particles. In this case, the shock is found to lie about half way to
the ejecta shell (similar to Figure \ref{fig:PostshockStripedWind}), so the
assumption $\delta R_{\rm neb} \sim R_{\rm ej}$ in calculating $\ell_c$
\citep[e.g.][]{Metzger2013a} appears to be well justified. However, unlike the
result shown in Figure \ref{fig:PostshockStripedWind} for Crab parameters, the
synchrotron frequency is $\gtrsim \unit[100]{MeV}$, and the efficiency
$\varepsilon_{\rm rad}$ in the forced reconnection zone is essentially
100\%. Accounting for the Klein-Nishina suppression, $\kappa \approx 10^{-2}$ at
$\unit[100]{MeV}$,
\begin{equation*}
  \tau_{\rm neb} = \kappa \, \sigma_T \, n_e \, \delta R_{\rm neb}
  \vspace{1 mm}
\end{equation*}
is found to be $\lesssim 10^{-2}$. Therefore, synchrotron photons produced at
the reverse shock traverse the nebula, and their fate is determined by the
albedo and optical depth of the shell. If insufficient radiation were reflected
back to the nebula, then $\varepsilon_{\rm rad}$ and thus $\ell_c$ would be
small, and the pair cascade could fail.

Of course, the preceding analysis ignores the enhancement of $n_e$ brought on by
a pair cascade. If one operates, then once again $\tau_{\rm neb} \gg 1$, photons
are trapped, radiation pressure forces the reverse shock to recede inwards, the
compactness parameter is kept large, and pair production continues. But this
illustrates the point made previously, that the pair cascade depends on itself
to survive, and is thus a complex problem. Unraveling the non-linear physics may
require a detailed calculation of fully coupled magnetic reconnection,
synchrotron radiation, photon propagation, and pair production processes. It is
difficult to say whether such a pursuit would be fruitful in the present
context. Nevertheless, similarly complex relativistic plasma systems were
examined by \cite{Timokhin2013}, and also by \cite{Beloborodov2016} with
interesting applications to pulsars and $\gamma$-ray burst prompt emission,
respectively. Continued research along these lines was advocated for by
\cite{Uzdensky2015}.

\acknowledgments J. Zrake acknowledges Yajie Yuan, Roger Blandford, Andrew
MacFadyen, Dan Kasen, Andrei Beloborodov, Brian Metzger, and Lorenzo Sironi for
inspiring discussions.

\bibliographystyle{apj}

\begin{appendix}

  \section{Turbulent relaxation model} \label{app:plasmoid-cascade}

  Coherent magnetic structures (eddies, or flux tubes) grow over time by merging
  with one another as a result of coalescence instability
  \citep[e.g.][]{Finn1977, East2015}, and they also grow or shrink due to the
  expansion or compression of comoving volume. For the latter reason it is
  convenient to utilize the eddy mass $m$ as a proxy for its scale. In three
  dimensions, magnetic structures are cylindrical flux tubes, having mass $m
  \sim \rho \ell^3$ (assuming their length and radius are comparable). The flux
  tubes are locally relaxed equilibria and thus generally helical
  \citep{Matthaeus1980, Matthaeus2012}, having comparable axial and azimuthal
  field strength. When net magnetic helicity is zero, left and right polarized
  flux tubes exist in equal number. Pairs can join by reconnecting their
  azimuthal field when their axial electric current vectors are parallel, but
  they only form a stable structure when their axial magnetic fields are also
  parallel; otherwise the axial field annihilates and renders the merged flux
  tube kink-unstable. Thus only half of the merging episodes (those occurring
  between like-polarized tubes) yield stable structures, and so the total number
  of eddies $s$ decreases by a factor of four in each stage of coalescence,
  $s_{n+1} = s_n / 4$. Stages proceed at the cascade rate
  \begin{equation*}
    \dot n = \frac{v_{\rm rec}}{\ell} \, ,
  \end{equation*}
  where $v_{\rm rec}$ is given roughly by the local Alfv{\'e}n speed $v_A =
  (\sigma / w)^{1/2}$. Each merging event conserves mass and magnetic helicity,
  so we have $m_{n+1} = 2 m_n$ and $h_{n+1} = 2 h_n$. Meanwhile, the magnetic
  energy per eddy drops according to $\epsilon_n = h_n / \ell_n$. The
  volume-averaged magnetic energy per unit mass is given by $\sigma_n = s_n
  \epsilon_n / \rho$. Here, the cascade number $n$ is a discrete version of the
  variable $\Delta$ used throughout the paper.

  When expansion is neglected ($\dot \rho = 0$), this heuristic yields a decay
  law in which the characteristic eddy scale increases over time as $\ell
  \propto t^{2/5}$, and the magnetic energy decays as $\sigma \propto
  t^{-6/5}$. Such behavior was seen by \cite{Zrake2014} in simulations of freely
  decaying, non-helical relativistic MHD turbulence in three
  dimensions. However, the heuristic's applicability to an expanding volume has
  not yet been explored directly with numerical simulations. Arguably, one
  expects the eddy mass to be a good proxy for its scale whenever the turbulence
  cascade rate $\dot n$ is fast compared with the secular evolution. Comoving
  volume in the pulsar wind expands at a rate $\omega_\perp = u / r$ in the
  transverse direction, and at $\omega_\parallel = \dot u / u$ in the
  longitudinal direction. If either of the expansion rates was faster than $\dot
  n$, the magnetic field pattern would become frozen into the flow, and get
  stretched in whichever direction expands faster. This situation corresponds to
  the magnetic free energy scale exceeding that of the local horizon. Provided
  the cascade is faster than both expansion rates, eddies maintain their
  isotropy in the comoving frame, and use of the mass coordinate in place of
  scale is well justified.

  \section{Plane-parallel reconnection picture}
  \label{app:reconnection-fronts}

  Here we review the conditions under which the plane-parallel reconnection
  picture of \cite{Lyubarsky2001} is equivalent to the turbulent relaxation
  model. Near the pulsar, the striped wind consists of cold, well-magnetized
  plasma. The azimuthal magnetic field alternates in direction every
  half-wavelength $\lambda = P_\star v / 2$. Magnetic reconnection operates
  around the reversals, causing slabs of hot plasma to expand at a speed $v_{\rm
    rec}$. The hot regions occupy a fraction $\Delta$ of the wind volume, and so
  $1 - \Delta$ is the surviving fraction of the wind's magnetic energy. Consider
  a cold plasma volume which is bounded by current layers centered at $r_1,
  r_2$. As the current layers expand, the one centered at $r_1$ extends forward
  to $r_+$, while the one centered at $r_2$ extends backward to $r_-$, so we
  have
  \begin{equation}
    \Delta = 1 - \frac{r_- - r_+}{r_2 - r_1} \, .
  \end{equation}
  Now, $\frac{dr_1}{dt} = v$ and to linear order $\frac{dr_2}{dt} = v + \lambda
  \frac{dv}{dr}$, where $\lambda = r_2 - r_1$ and $t$ here denotes time in the
  pulsar frame. The velocity gradient is given by $\frac{dv}{dr} =
  \omega_\parallel \gamma^{-3}$ where $\omega_\parallel = \dot u / u =
  \frac{du}{dr}$ is the acceleration rate. The reconnection fronts advance
  according to the relativistic velocity addition of the flow and $v_{\rm rec}$,
  \begin{equation}
    \frac{d}{dt} r_\pm = \frac{v_\pm \pm v_{\rm rec}}{1 + v_\pm v_{\rm rec}} \, ,
  \end{equation}
  where $v_\pm = v + (r_\pm - r_1) \frac{dv}{dr}$. We also assume that $r_1 +
  r_2 = r_- + r_+$. Since the wind is ultra-relativistic, we expand $\dot \Delta
  = \gamma \frac{d \Delta}{dt}$ in powers of $\gamma^{-1}$,
  \begin{equation}
    \dot \Delta = \frac{2 v_{\rm rec}}{1 - v_{\rm rec}^2} \left\{
    \frac{1}{\lambda} \gamma^{-1} - \omega_\parallel \left[1 + v_{\rm rec}(1 -
      \Delta) \right] \gamma^{-2} + \mathcal{O}(\gamma^{-3})\right\} \, .
    \label{eqn:delta-dot-full}
  \end{equation}
  Further dropping terms of order higher than $v_{\rm rec}^2$, we arrive at
  \begin{equation*}
    \dot \Delta = \frac{2 v_{\rm rec}}{\lambda \gamma} + \mathcal{O}(v_{\rm rec}^2) +
    \mathcal{O}(\gamma^{-2}) \, ,
  \end{equation*}
  which is the same as Equation \ref{eqn:reconnection-front-speed}, and
  equivalent to Equation C3 of \cite{Kirk2003}. The second order term in
  Equation \ref{eqn:delta-dot-full} inhibits the progress of reconnection fronts
  by stretching the background flow, but remains small provided $\lambda \gg
  h_{\rm rec}$, where $h_{\rm rec} = u_{\rm rec} / \omega_\parallel$ is the
  local horizon scale with respect to the reconnection speed. Sticking to first
  order in $\gamma^{-1}$ is thus sufficiently accurate unless reconnection were
  somehow to proceed ultra-relativistically.

  \section{Shock jump conditions}
  \label{app:kc-jump=conditions}
  The Kennel-Coroniti jump conditions are given by conservation of mass, magnetic
  flux, energy, and momentum across the shock,
  \begin{eqnarray*}
    u_1 \rho_1 &=& u_2 \rho_2 \\ u_1 \sigma_1 &=& u_2 \sigma_2 \\ \gamma_1 w_1 &=&
    \gamma_2 w_2 \\ u_1 w_1 + \frac{1}{u_1}\left(\frac{p_1}{\rho_1} +
    \frac{\sigma_1}{2}\right) &=& u_2 w_2 + \frac{1}{u_2}\left(\frac{p_2}{\rho_2}
    + \frac{\sigma_2}{2}\right) \, ,
  \end{eqnarray*}
  where subscripts 1 and 2 refer to values just ahead of, and just behind the
  shock respectively. Together with the $\Gamma$-law equation of state, these
  yield the following equation for the density jump $\delta \equiv \rho_1 /
  \rho_2$ across the shock,
  \begin{equation}
    \eta \left(\delta \frac{1 + \frac{\Gamma}{\Gamma - 1} u_1^2}{\sqrt{1 + u_1^2}}
    - \frac{1 + \frac{\Gamma}{\Gamma - 1} u_1^2 \delta^2}{\sqrt{1 + u_1^2
        \delta^2}} \right) - \sigma_1 \left( \frac{\Gamma - 2}{\Gamma - 1} \right)
    \left( \frac{\delta^2 - 1}{2 \delta} \right) = \delta - 1 \, .
    \label{eqn:kc-jump-conditions}
  \end{equation}

\end{appendix}
\end{document}